\documentclass[A4,paper,12pt]{article}

\usepackage{amsmath}
\usepackage{amssymb}
\usepackage[cp1250]{inputenc}
\usepackage{placeins}
\usepackage{graphicx}

\newcommand{\pd}{\partial}
\renewcommand{\d}{\mathsf{d}}
\newcommand{\opr}[1]{\mathsf{#1}}
\newcommand{\form}[1]{\mathsf{#1}}

\newcommand{\laplace}{\triangle}
\newcommand{\grad}{\bigtriangledown}
\renewcommand{\O}{\mathcal{O}}

\newcommand{\man}[1]{\mathcal{#1}}
\renewcommand{\r}{\rho(\theta,\,\phi)}
\newcommand{\e}{\epsilon}

\newenvironment{proof}{{\sl Proof:}\rm}{$\blacksquare$ \\ \smallskip}

\newtheorem{definition}{Definition}[section]

\newtheorem{theorem}[definition]{Theorem}
\newtheorem{remark}[definition]{Remark}
\newtheorem{proposition}[definition]{Proposition}

\begin{document}

\title{On geometric perturbations of critical Schr\"odinger
operators with a surface interaction}

\author{Pavel Exner$^{1,2}$ and Martin Fraas$^{3}$}
\date{\small $^{1}$Nuclear Physics Institute, Czech Academy of Sciences,
25068 \v{R}e\v{z} near Prague \\
$^{2}$Doppler Institute, B\v{r}ehov\'{a} 7, 11519 Prague,
Czechia \\  $^{3}$Physics Department, Technion, Haifa 32000, Israel \\
\emph{exner@ujf.cas.cz, fraas@ujf.cas.cz} }

\maketitle

%%%%%%%%%%%%%%%%%%%%%%%%%%%%%%%%%%%%%%%%%%%%%%%%%%
\begin{abstract}
\noindent We study singular Schr\"odinger operators with an
attractive interaction supported by a closed smooth surface $\man
A\subset\mathbb{R}^3$ and analyze their behavior in the vicinity
of the critical situation where such an operator has empty
discrete spectrum and a threshold resonance. In particular, we
show that if $\man A$ is a sphere and the critical coupling is
constant over it, any sufficiently small smooth area preserving
radial deformation
gives rise to isolated eigenvalues. On the other hand, the
discrete spectrum may be empty for general deformations. We also
derive a related inequality for capacities associated with such
surfaces.
\end{abstract}

\setcounter{equation}{0}
%%%%%%%%%%%%%%%%%%%%%%%%%%%%%%%%%%%%%%%%%%%%%%%%%%
\section{Introduction}
\label{sec:intro}

Relations between geometrical and spectral properties belong to
traditional questions in mathematical physics. Recently an
isoperimetric inequality was derived for two-dimensional
Schr\"odinger operators with a singular attractive interaction
supported by a closed loop \cite{ex, ehl}. It claims that if the
coupling is constant along the loop the ground state eigenvalue is
maximized in the class of all loops of a fixed length by a circle.
The result has interested connections to both the classical
electrodynamics \cite{ex, acfgh} and a class of isoperimetric
inequalities of a purely geometric nature \cite{acfgh, luko, efh}.

One can ask naturally whether the result has a higher-dimensional
counterpart, that is, for Schr\"odinger operators with an
attractive interaction supported by a closed hypersurface. Our aim
in this paper is to address this question in dimension three. It
is clear from the outset that such a problem is more involved. One
difference comes from the fact that the two-dimensional operator
with an attractive interaction has always a non-empty discrete
spectrum while its three-dimensional counterpart may be positive
if the coupling is sufficiently weak. This opens a possibility, on
the other hand, that such operator may or may not have bound
states depending on geometric perturbations of the interaction
support.

To be specific, choose a a closed smooth surface $\man
A\subset\mathbb{R}^3$ and consider the operator formally given as
$-\Delta-\alpha\delta(x-\man A)$ with an attractive coupling
constant over $\man A$. If we choose $\alpha>0$ such that for a
spherical $\man A$ the operator is critical, i.e. any larger
$\alpha$ will produce a nontrivial discrete spectrum, one might
expect in analogy with the two dimensional case that any
deformation of $\man A$ preserving its area would lead to
occurrence of negative eigenvalues. In reality the situation is
more complicated. We are going to show that the above claim is
valid for small enough smooth radial deformations of $\man A$,
however, it fails generally: there are ``large'' area-preserving
deformations which leave the operator positive.

On the other hand, one can prove another global result for
deformations of $\man A$. It exhibits again a relation to
electrostatics, although it is now different from the one
mentioned above: the preserved quantity to replace the loop length
of the two-dimensional situation is not the area of $\man A$ but
the \emph{capacity} of the capacitor represented by the surface.

Let us briefly described the contents of the paper. After
presenting in the Sections \ref{sec:prelim} and \ref{sec:crit} the
necessary preliminaries, we shall state in
Section~\ref{sec:result} the main results of this paper. Their
proofs and discussion will follow. In particular, the proof of
Theorem \ref{def:capacity} is given in the section
\ref{sec:minimax}, and the section \ref{sec:local} deals with
local deformations of the sphere which are the contents of Theorem
\ref{def:local}. Finally, in Section \ref{sec:global} we will show
that the claim of Theorem \ref{def:local} cannot be extended to
general deformations.

\setcounter{equation}{0}
%%%%%%%%%%%%%%%%%%%%%%%%%%%%%%%%%%%%%%%%%%%%%%%%%%
\section{Singular interactions on a surface}
\label{sec:prelim}

Let $\man A$ be a closed smooth surface in $\mathbb{R}^3$ and
$\nu$ the ``natural'' measure on $\man A$ induced by its embedding
into the Euclidean space, in other words, Lebesgue measure in the
appropriate local charts of $\man A$. Consider next a bounded
Borel measurable function $\alpha(x) : \man A \to \mathbb{R}$ and
$\phi,\,\psi \in \mathcal{W}^{2,1}(\mathbb{R}^3)$ and define the
quadratic form
 % ------------- %
$$
\form{Q}_\alpha(\phi,\,\psi) := \int_{\mathbb{R}^3} \grad \phi(x)
\cdot \overline{\grad \psi(x)}\, \d x\, - \int_{\man A} \alpha(x)
\phi(x) \overline{\psi(x)}\, \d \nu(x)\,;
$$
 % ------------- %
by applying the Green formula one has
 % ------------- %
\begin{multline}
\form Q_\alpha(\phi,\,\psi) = \int_{\mathbb{R}^3} -(\laplace
\phi)(x) \overline{\psi(x)}\,\d x \, - \int_{\man A}
\left(\frac{\pd \phi(x)}{\pd n_e} + \frac{\pd \phi(x)}{\pd
n_i}\right) \overline{\psi(x)}\,\d \nu(x)\\
- \int_{\man A} \alpha(x) \phi(x) \overline{\psi(x)}\, \d
\nu(x)\,, \nonumber
\end{multline}
 % ------------- %
where $n_e,\,n_i$ are exterior and interior normal, respectively.
This suggest that the operator
 % ------------- %
$$ \opr{H}_\alpha := -\laplace\,, $$
 % ------------- %
defined on functions which are locally $\mathcal{W}^{2,1}$ away
from $\man{A}$ and satisfy
 % ------------- %
\begin{equation}
\frac{\pd \phi(x)}{\pd n_e} + \frac{\pd \phi(x)}{\pd n_i} =
-\alpha(x) \phi(x)\,,
  \label{BoundCon}
\end{equation}
 % ------------- %
for every $x \in \man{A}$ is self-adjoint and corresponds to the
form $\form Q_\alpha$; for a proper justification of this claim
and related results see \cite{beks, pos}. The operator
$\opr{H}_\alpha$ is naturally interpreted as a singular
Schr\"{o}dinger operator with the interaction supported by the
surface $\man A$ and the coupling ``constant'' $\alpha(x)$. It
motivates us to define the global strength of the interaction,
 % ------------- %
\begin{equation}
[\alpha]:=\int_\man{A} \alpha(x)\,\d \nu(x) \label{strength}
\end{equation}
 % ------------- %
and the relative density of the interaction,
 % ------------- %
\begin{equation}
\hat\alpha(x):= \frac{\alpha(x)}{[\alpha]}\,. \label{density}
\end{equation}
 % ------------- %
The last definition makes sense, of course, only if $[\alpha]\ne
0$, which will be true in our case, since we are going to consider
only singular Schr\"{o}dinger operator with attractive
interactions, $\alpha(x)>0$ for all $x \in \man{A}$.

A situation of particular interest is the case of a function
$\alpha$ constant over the surface $\man{A}$ taking a value
$\alpha_0>0$ there; for such a function we obviously have
$[\alpha]= \alpha_0 S$ and $\hat\alpha(x) = S^{-1}$, where
$S:=\nu(\man{A})$ is the surface area.

\setcounter{equation}{0}
%%%%%%%%%%%%%%%%%%%%%%%%%%%%%%%%%%%%%%%%%%%%%%%%%%
\section{Criticality}
\label{sec:crit}

Let us first recall the standard criticality notions. If
$\opr{H}:=-\laplace + V(x) $ is a Schr\"{o}dinger operator and
$W(x) \geq 0$ an arbitrary nonzero and compactly supported
function, we say that $\opr{H}$ is \emph{subcritical} if for
$\epsilon > 0$ small enough the operator $-\laplace +V(x) -
\epsilon W(x)$ remains to be positive. Correspondingly, $\opr{H}$
is \emph{critical} if its positivity depends on the sign of the
perturbation, and \emph{supercritical} if $-\laplace +V(x) +
\epsilon W(x)$ is negative for sufficiently small $\epsilon > 0$.
Clearly, the interaction is supercritical if and only if the
operator $\opr{H}$ has a negative bound state.

These notions can be carried over directly to the singular case
where one has to consider the operators $\opr{H}:=-\laplace \pm
\epsilon W(x)$ with the boundary conditions (\ref{BoundCon}) as
perturbations of $\opr{H}_\alpha$. Let us collect without proofs
several simple facts about criticality. Recall that the solution $u$
of $\opr{H} u =0$ is said to have minimal growth at infinity if
for any other solution $v$ there is a constant $C$ such that $C v(x)
>  u(x)$ holds for all $|x|$ sufficiently large \cite{Pinchover}.
With this notion, we can state the following result.
 % ------------- %
\begin{proposition}
\label{def:criticality} Let $\opr{H}_\alpha$ be the singular
Schr\"{o}dinger operator $\opr{H}_\alpha$ with the interaction
supported by $\man{A}$ as described above. Then the following
claims are equivalent:
\begin{itemize}
 \setlength{\itemsep}{-3pt}
 \item[(i)] $\opr{H}_\alpha$ is critical
 \item[(ii)] For any bounded function $\beta>0$, $\opr{H}_{\alpha
   \pm \beta}$ is supercritical or subcritical respectively.
 \item[(iii)] The equation $\opr{H}_\alpha u = 0$ has a positive
   solution with a minimal growth at infinity.
\end{itemize}
\end{proposition}
 % ------------- %
In the following we will always assume that the relative
interaction density of the operator $\opr{H}_\alpha$ is fixed and
discuss how the properties of the operator depend on $[\alpha]$.
Since $\opr{H}_0 = -\laplace$ is subcritical in dimension three it
is clear that $\opr{H}_\alpha$ is subcritical for $[\alpha]$ small
enough and supercritical for $[\alpha]$ large. Our main concern is
the value at which $\opr{H}_\alpha$ is critical; we will denote as
$[\hat\alpha]_c$.

Relations between $[\hat\alpha]_c$ and the geometric properties of
the interaction are of a natural interest. In particular, one can
ask about the critical strength for a given surface and a fixed
interaction density, and about the dependence of the critical
strength on the shape of the surface. To get some insight into
these questions, we are going to compare the critical strength
$[\hat\alpha]_c$ with the surface area $S$ and its capacity $C$.

A comparison requires to select appropriate quantities. As an
inspiration, note that in the particular situation when the
surface $A$ is a sphere of radius $R$ and $\alpha$ is constant
over it, $\hat\alpha(x)=(4\pi R^2)^{-1}$ for any $x\in\man A$, one
can compute the above named quantities explicitly,
 % ------------- %
$$
[\hat\alpha]_c=4 \pi R\,, \quad S=4 \pi R^2\,,\quad C=R\,.
$$
 % ------------- %
With this example on mind we define the \emph{interaction radius}
$\overline{[\hat\alpha]}_c:=[\hat\alpha]_c/(4 \pi)$ and the
classical \emph{surface radius} $\overline{S} := \sqrt{S/(4
\pi)}$. Note that while $\alpha$ has, physically speaking,
dimension $\mathrm{length}^{-1}$, the integral quantities
$[\alpha]$ and $[\hat\alpha]_c$ has the dimension of length; it
will be of importance in the following that the quantities
$\overline{[\hat\alpha]}_c$ and $\overline{S}$ scales in the same
manner.

It is well known \cite{PolyaSzego} that $\overline{S}$ and $C$ are
\emph{not} comparable, in particular, that $\overline{S} \geq C$
need not hold even in the vicinity of sphere. Our aim is to try to
compare these two quantities to the interaction radius for a fixed
relative interaction density. Recall that our original motivation
to study this problem mentioned in the introduction was to find
out whether any surface-preserving deformation of a critical
sphere with the interaction $\alpha(x)=\alpha_0$ of the constant
relative density $S^{-1}$ will produce a bound state. In terms of
the notions introduced above we can say that it will be true if
$\overline{S} > \overline{[\hat\alpha]}_c$ holds for the deformed
surface where $\overline{S}$ is the radius of the critical sphere.

\setcounter{equation}{0}
%%%%%%%%%%%%%%%%%%%%%%%%%%%%%%%%%%%%%%%%%%%%%%%%%%
\section{Main results}
\label{sec:result}

Let us formulate now the claims we are going to demonstrate. First
of all, the above mentioned property will be valid if the quantity
preserved at the deformation is not the surface area but the
capacity. To this aim we can think of the surface $\man A$ as of a
capacitor charged with an unit charge denoting by $\sigma(x)$ the
corresponding charge density,
 % ------------- %
 \begin{equation} \label{unitch}
\int_{\man A} \sigma(x) \d \nu(x) = 1.
 \end{equation}
 % ------------- %
For the definition of capacity, charge density and their
properties, see for instance \cite[Chap.~II]{PolyaSzego} and also
a brief recapitulation in the next section.

\begin{theorem}
\label{def:capacity} Let $\man A$ be a surface with the capacity
$C$, then
\begin{itemize}
 \item[(a)] The operator $\opr{H}_{\alpha}$ is critical if $\alpha=4
 \pi C \sigma$, i.e. $\overline{[\hat\sigma]}_c=C$.
 \item[(b)] Let $\opr{H}_\alpha$ correspond to a constant interaction,
 $\alpha(x)=\alpha_0$ for all $x\in{\man A}$, then $\overline{[\hat\alpha]}_c \leq
C$.
\end{itemize}
\end{theorem}
 % ------------- %

\begin{remark}
{\rm If $\man{A}$ is a sphere, then the equality $C = \overline{[\hat\sigma]}_c$ holds in the part (b) of the above theorem, while the proof of the converse statement is an open question.
Note that it is equivalent to the claim that the sphere is the only closed capacitor with a constant surface charge density. It is a common knowledge that the charge tends to concentrate at the points of high curvature, but we have not found a rigorous elaboration of this assertion.}

\end{remark}

In the surface-preserving situation we have a weaker result valid
for a class of gentle deformations. Let $A$ be a surface defined
by the equation
 % ------------- %
$$
r \equiv r(\theta,\,\phi) = r_0 \left(1 +
\epsilon\rho(\theta,\,\phi) \right),
$$
 % ------------- %
where $r,\,\theta,\,\phi$ are the spherical coordinates, $\rho$ is
a fixed smooth and nonzero function on the unit sphere, and
$\epsilon \in (0,\|\rho\|_\infty^{-1})$. We will speak about
smooth \emph{radial} deformations of the sphere.
 % ------------- %
\begin{theorem}
\label{def:local} Let $A$ be a surface described above and let $S$
denote its area. For the corresponding operator $\opr{H}_\alpha$
with a constant interaction, $\alpha(x)=\alpha_0$ for all
$x\in{\man A}$, we have
 % ------------- %
\begin{equation}
\label{motak} \overline{S} >  \overline{[\hat\alpha]}_c
\end{equation}
 % ------------- %
provided $\epsilon$ is small enough.
\end{theorem}
 % ------------- %
In contrast to the previous result this claim cannot be extended
to general deformations. In Sec.~\ref{sec:global} we will provide
an example of an area-preserving deformation for which the
inequality (\ref{motak}) is violated.

Notice also that the previous result does not help us here. The
part (b) of Theorem~\ref{def:capacity} would imply
Theorem~\ref{def:local} if $\overline{S} \geq C$ held in the
vicinity of the sphere, however, we have mentioned already that
this is not the case.

\setcounter{equation}{0}
%%%%%%%%%%%%%%%%%%%%%%%%%%%%%%%%%%%%%%%%%%%%%%%%%%
\section{Capacity and Gauss variational principle}
\label{sec:electrostatic}

Let $\man A$ be a closed smooth surface, which is regard here as a
capacitor. It is well know that up to multiplicative constant
there is a unique solution of the Laplace equation
 % ------------- %
$$
\laplace u = 0\,,
$$
 % ------------- %
in the exterior of $\man A$, denoted by $\man A^\mathrm{ext}$,
such that $u$ is constant on the surface and zero at infinity.
This solution describes the potential and its negative normal
derivative at the surface is the charge density,
 % ------------- %
$$
\sigma(x) := -\frac{1}{4 \pi}\frac{\pd u(x)}{\pd n_e}\,,\quad x
\in \man A\,.
$$
 % ------------- %
If we normalize the potential to the unit charge on the capacitor
${\man A}$ as expressed  by the relation (\ref{unitch}), the
capacity $C$ is related to the value on the surface by
 % ------------- %
\begin{equation}
u(x) = C^{-1}\,,\quad x \in \man A\,.
\label{eq:cap}
\end{equation}
 % ------------- %
If we expand the potential $u$ into spherical harmonics,
 % ------------- %
\begin{equation}
\label{PotRozvoj} u(x) =  \frac{S_0}{r} + \sum_{n=1}^\infty
S_n(\theta,\,\phi)\, r^{-n-1}\,,
\nonumber
\end{equation}
 % ------------- %
where $S_n(\theta,\,\phi) = \sum_{m=-n}^n c_m
Y_{mn}(\theta,\,\phi)$ with some coefficients $\{c_m\}$ is the
contribution of order $n$, then by Green's theorem we infer that
$S_0 = C^{-1}$. Moreover, by the well-known formula for the Green
function one gets
 % ------------- %
\begin{equation}
 u(x) = \int_\man A \frac{\sigma(y)}{|x-y|}\, \d
\nu(y)\, \nonumber
\end{equation}
 % ------------- %
 and combining with the (\ref{eq:cap}) we get that for $x \in \man A$
 %------------
\begin{equation}
 \label{PotGreen}
 \int_\man A \frac{\sigma(y)}{|x-y|}\, \d
 \nu(y) = \frac{1}{C}.
\end{equation}
%-----------
We will also need one more characterization of the capacity,
called Gauss variational principle
\cite[Chap.~II.2.8]{PolyaSzego}: let $\man A$ be a capacitor and
$\mu(x)$ a positive measurable function on $A$ satisfying
 % ------------- %
$$
\int_\man A \mu(x) \d \nu(x) = 1\,,
$$
 % ------------- %
then it holds
 % ------------- %
\begin{equation}
\label{Gauss} \int_{\man A \times \man A} \frac{\mu(x)
\mu(y)}{|x-y|}\, \d \nu(x) \d \nu(y) \geq \frac{1}{C}
\end{equation}
 % ------------- %
and the equality holds in this relation if and only if $\mu(x) =
\sigma(x)$ -- cf. (\ref{PotGreen}).

\setcounter{equation}{0}
%%%%%%%%%%%%%%%%%%%%%%%%%%%%%%%%%%%%%%%%%%%%%%%%%%
\section{Krein-like resolvent formula}
\label{sec:minimax}

As usual the best way to analyze spectral properties is to employ
the resolvent. For singular Schr\"odinger operators it is made
possible due to the existence of an explicit resolvent formula of
Krein (or Birman-Schwinger) type. A detailed discussion can be
found in \cite{beks, pos}, here we limit ourselves to quoting a
few simple facts. Given $z \in \mathbb{C}\setminus(0,\infty)$ we
use the free resolvent with the kernel
 % ------------- %%%
$$
G(z)(x,y):=\frac{1}{4 \pi} \frac{\exp(i \sqrt{z} |x -y|)}{|x-y|}
$$
 % ------------- %%
to define a pair of operators,
 % ------------- %
$$
\opr{G}(z) : L^2(\man A,\d\nu) \to L^2(\mathbb{R}^3)\,, \quad
\opr{G}(z)u(x):=\int_\man{A} G(z)(x,\,y) u(y)\, \d\nu(y)
$$
 % ------------- %
and
%%%%
$$
\opr{\Gamma}(z) : L^2(\man A,\d\nu) \to L^2(\man A,\d\nu)\,, \quad
\opr{\Gamma}(z)u(x):=\int_\man{A} G(z)(x,\,y) u(y)\, \d\nu(y)\,.
$$
 % ------------- %
Then the indicated Krein-type resolvent formula reads
 % ------------- %
$$
(\opr{H}_\alpha(x) -z)^{-1} = (\opr{H}_0 -z)^{-1} + \opr{G}(z)(I -
\alpha \opr{\Gamma}(z))^{-1} \alpha \opr{G}(z)^*,
$$
where $\alpha$ and  $I$ are the multiplication operator by
$\alpha(\cdot)$ and the unit operator on $L^2(\man A,\d\nu)$,
respectively. Moreover, it yields a simple characterization of the
point spectrum which generalizes to the singular case the
Birman-Schwinger principle: $-\kappa^2$ is an eigenvalue if and
only if the operator
 % ------------- %
$$
I - \alpha \opr{\Gamma}(i\kappa)
$$
 % ------------- %
has a nontrivial kernel, and its dimension coincides with the
eigenvalue multiplicity. This suggests that $\opr{H}_\alpha$ is
critical if $I - \alpha \opr{\Gamma}(0)$ has a nontrivial kernel.
We make this claim more precise in the following proposition.
 % ------------- %
\begin{proposition}
\label{def:crucial}
Let $\hat\alpha(\cdot)$ be a relative interaction density on the
surface $\man{A}$ and $\Gamma\equiv\Gamma(0)$ the above defined
operator with the kernel
 % ------------- %
$$
\Gamma(x,y)=\frac{1}{4 \pi}\frac{1}{|x-y|}\,, \quad
x,y\in\man{A}\,,
$$
 % ------------- %
then the critical strength is given by $[\hat\alpha]_c =
\|\hat\alpha \Gamma\|^{-1}$.
\end{proposition}
 % ------------- %
\begin{proof}
First we check that for any measurable, bounded and positive
$\hat\alpha(\cdot)$, the number $\|\hat\alpha \opr{\Gamma}\|$ is
an eigenvalue (naturally, the largest one) of the operator
 % ------------- %
$$\hat\alpha \opr{\Gamma}\:: \quad L^2(\man A,\d\nu)
\to L^2(\man A,\d\nu)$$
 % ------------- %
and the corresponding eigenfunction is positive. To see the first
part of the statement, observe that the operator $\opr{G} \equiv
\opr{G}(0)$ maps $L^2(\man A,\d\nu)$ to the set of solutions of
the equation $-\laplace u = 0$ away from $\man A$. This means, in
particular, that $\mathrm{Ran}\,\opr{G} \subset
W^{2,\,2}_\mathrm{loc}(\mathbb{R}^3 \setminus \man A)$ and by the
well known properties of trace operator \cite[Section VI.4]{stein}
we have $\mathrm{Ran}\,\Gamma \subset W^{1,2}(\man A,\d\nu)$ which
is compactly embedded into $ L^2(\man A,\d\nu)$. This shows that
$\opr{\Gamma}$ is a compact operator, thus $\hat \alpha
\opr{\Gamma}$ is compact as well and the upper edge of its
spectrum, $\|\hat \alpha \opr{\Gamma}\|$, is an eigenvalue.
Furthermore, the operator $\hat \alpha \opr{\Gamma}$ is positivity
improving, hence the positivity of the corresponding eigenfunction
follows in the standard way -- cf.~\cite[Thm~XII.44]{RS}.

Consequently, the operator $\alpha_c \opr{\Gamma}:=
\frac{1}{\|\hat \alpha \opr \Gamma\|} \hat\alpha \opr{\Gamma}$
corresponding to a given relative density $\hat \alpha(x)$  has
the number one as its largest eigenvalue, and a positive
eigenfunction $\phi$ is associated with it. We put
 % ------------- %%
$$
u(x):=(\opr{G}\phi)(x)\,,
$$
 % ------------- %
then the explicit form of the kernel yields for any $x \in \man A$
 % ------------- %
\begin{eqnarray*}
\frac{\pd u(x)}{\pd n_+} + \frac{\pd u(x)}{\pd n_-} &=& -\phi(x) \\
u(x) = (\tau \opr{G} \phi)(x) = (\opr{\Gamma} \phi)(x) &=&
\alpha_c(x)^{-1} \phi(x) \,,
\end{eqnarray*}
 % ------------- %
so the above function $u$ belongs to $D(\opr{H}_{\alpha_c})$ being
a positive solution to $\opr{H}_{\alpha_c(x)} u  = 0$. Moreover,
the asymptotic of $\opr{G}(x,y)$ shows that $u(x) \sim
const\,|x|^{-1}$ at large distances being thus of the minimal
growth at infinity. Proposition~\ref{def:criticality} then implies
that $\opr{H}_{\alpha_c}$ is critical and $[\hat\alpha]_c =
\|\hat\alpha \Gamma\|^{-1}$ what we have set out to prove.
\end{proof}

\noindent With this preliminary, the \emph{proof of theorem
\ref{def:capacity}} follows easily: \\ [.2em]
(a) Using the relation (\ref{PotGreen}) we get
 % ------------- %%
$$
\frac{1}{4 \pi}\, \sigma(x) \int \frac{\sigma(y)}{|x-y|}\,
\d\nu(y) = \frac{1}{4 \pi C}\, \sigma(x)
$$
 % ------------- %%
for $x\in\man A$, and since $\sigma(x)$ is positive everywhere we
conclude that $1/(4 \pi C)$ is the maximal eigenvalue of $\sigma
\opr{\Gamma}$, and hence its norm. Consequently, we have
$$
\overline{[\sigma]}_c:=C\,.
$$
(b) By a simple variational estimate we have that for any unit vector
$\psi$
 % ------------- %
$$
\|\hat \alpha \opr{\Gamma}\| \geq \int_\man{A} \overline{\psi(x)}
\hat \alpha(x) (\Gamma \psi)(x)\, \d\nu(x)
$$
 % ------------- %
and the equality holds if and only if $ \psi$ corresponds to the
maximal eigenvalue. For the constant relative density $\hat\alpha
:= S^{-1}$ and $\psi = S^{-1/2}$ we can estimate the right-hand
side using the Gauss variational principle,
 % ------------- %
$$
\frac{\hat\alpha}{4 \pi} \int_{\man A \times \man A}
\frac{\overline{\psi(x)} \psi(y)}{|x-y|}\, \d \nu(x) \d \nu(y) =
\frac{1}{4 \pi} \int_{\man A \times \man A} \frac{S^{-2}}{|x-y|}\,
\d \nu(x) \d \nu(y) \geq \frac{1}{4 \pi}\frac{1}{C},
$$
 % ------------- %
which gives $\|\hat \alpha \opr{\Gamma}\|^{-1}\leq 4 \pi C$, or
equivalently, $\overline{[\hat\alpha]}_c \leq C$.

\setcounter{equation}{0}
%%%%%%%%%%%%%%%%%%%%%%%%%%%%%%%%%%%%%%%%%%%%%%%%%%
\section{Local deformations of a sphere}
\label{sec:local}

In this section we will work in the spherical coordinates
$(r,\,\theta,\,\phi)$. We start from a sphere; due to the natural
scaling properties it is sufficient to consider the case of unit
radius. It is straightforward to check that for such a sphere the
critical interaction constant over the surface is $\alpha(x) = 1$
for all $x\in\man A$, and the corresponding positive solution with
minimal growth at infinity is $u = r^{-1}$ in the exterior of the
sphere (denoted as $\man A^\mathrm{ext}$) and $u=1$ inside.

We consider a radially deformed surface $\man A_\varepsilon$
defined by the equation
 % ------------- %
$$
r(\theta,\,\phi)=1 + \epsilon \rho(\theta,\,\phi),
$$
 % ------------- %
where $\rho$ is a fixed smooth and nonzero function on the unit
sphere, and $\epsilon \in (0,\|\rho\|_\infty^{-1})$, in
particular, $\man A_0$ is the sphere mentioned above. Our aim is
to find a perturbed solution $u_\epsilon$ and the constant
interaction, $\alpha(x) =\alpha_\epsilon$ for all $x\in\man A$,
such that the solution will remain positive and bounded, in other
words, $\alpha_\epsilon$ would be the critical strength for the
said constant singular interaction on the surface $\man
A_\epsilon$. We are going to show that for a nontrivial
$\rho(\theta,\,\phi)$ the inequality $\overline{S}_\epsilon
\alpha_\epsilon < 1$ is valid for small and nonzero $\epsilon$,
where $\overline{S}_\epsilon$ is the corresponding surface radius.
This will prove Theorem~\ref{def:local}, because $[\hat\alpha]_c=
\alpha_\epsilon S_\epsilon$ would then yield
$\overline{[\hat\alpha]}_c = \overline{S}_\e^2 \alpha_\e <
\overline{S}_\e$ for such an $\epsilon$.

As usual in such situations the method is to employ asymptotic
expansion in powers of~$\epsilon$. We put
 % ------------- %%
$$\rho(\theta,\,\phi) = \sum_n X_n(\theta,\,\phi),$$
 % ------------- %
where $X_n$ is a spherical harmonic of order $n$ and we will look
for a solution $u_\epsilon$ of $-\laplace u = 0$ away from the
surface, such that it is continuous on $\man A_\epsilon$ and has
the corresponding jump in normal derivative there, i.e.
 % ------------- %
\begin{equation}
\frac{\pd u_\epsilon(x)}{\pd n_e} + \frac{\pd u_\epsilon(x)}{\pd
n_i} = - \alpha_\epsilon u_\epsilon(x) \label{ebc}
\end{equation}
 % ------------- %
holds for $x \in \man A_\epsilon$. We will seek it in the form
 % ------------- %
\begin{eqnarray}
u^\mathrm{ext}_\epsilon(r,\theta,\,\phi) &=& \frac{1}{r} +
\epsilon \sum_{n=1}^{\infty} S_n^{(1)}(\theta,\,\phi) r^{-n-1}
+\epsilon^2 \sum_{n=1}^{\infty} S_n^{(2)}(\theta,\,\phi) r^{-n-1}
+ \O(\epsilon^3)\,, \nonumber \\
u^\mathrm{in}_\epsilon(r,\theta,\,\phi) &=& 1 + \epsilon
\sum_{n=0}^{\infty} R_n^{(1)}(\theta,\,\phi) r^{n} +\epsilon^2
\sum_{n=0}^{\infty} R_n^{(2)}(\theta,\,\phi) r^{n}
+\O(\epsilon^3)\,, \nonumber \\
\alpha_\epsilon &=& 1 +\epsilon \alpha^{(1)} + \epsilon^2
\alpha^{(2)} + \O(\epsilon^3)\,, \nonumber
\end{eqnarray}
 % ------------- %
where $S_n^{(i)},\,R_n^{(i)}$ are spherical harmonics of order
$n$. Such an Ansatz will guarantee that
$u_\e^\mathrm{ext}\,(u_\e^\mathrm{in})$ is bounded solution of
$-\laplace u =0$ in the exterior (respectively, interior) of the
surface, and furthermore, that $u_\e^\mathrm{ext}$ has the $1/r$
asymptotics at infinity. It is convenient to allow the summation
in the definition of $u^\mathrm{ext}_\e$ to run also from zero by
putting $S_0^{(i)}(\theta,\,\phi) = 0$; this will allow us to
write some formul{\ae} below in a more compact form.

Due to the nature of the deformation each element of $\man
A_\epsilon$ can be uniquely characterized by $x=(\theta,\,\phi)$.
The corresponding surface element equals
 % ------------- %
\begin{multline}
\d \nu_\epsilon(\theta,\,\phi) = \bigg\{1 + 2 \epsilon
\rho(\theta,\,\phi)+ \epsilon^2 \rho(\theta,\,\phi)^2 \\ +
\frac{1}{2} \e^2 \bigg( \left(\frac{\pd \rho(\theta,\,\phi)}{\pd
\phi}\right)^2 + \frac{1}{\sin^2 \phi} \left(\frac{\pd
\rho(\theta,\,\phi)}{\pd \theta}\right)^2 \bigg) \bigg\} \sin \phi
\, \d \theta\, \d \phi + \O(\epsilon^3)\,, \label{surel}
\end{multline}
 % ------------- %
and for the exterior normal vector $n_\epsilon$ to $\man
A_\epsilon$ we find
 % ------------- %
\begin{multline}
\d \nu_\epsilon(\theta,\,\phi) n_\epsilon(\theta,\,\phi) =
(1+\epsilon^2 \r^2)\, \hat r\, \cos \phi  \\ - \epsilon(1
+\epsilon \r) \frac{\pd \r}{\pd \phi}\, \hat \phi\, \sin \phi - (1
+ \e \r) \frac{\pd \r}{\pd \theta}\, \hat \theta\,, \nonumber
\end{multline}
 % ------------- %
where we have introduced the standard unit vector triple $(\hat
r,\,\hat \theta,\,\hat \phi)$ at the surface point characterized
by $x$. Expanding $1/r$ we find
 % ------------- %
\begin{eqnarray}
u_\epsilon^\mathrm{ext}(x) &=& 1
-\e \sum_{n=0}^\infty X_n(\theta,\,\phi)
+ \e \sum_{n=0}^\infty S_n^{(1)}(\theta,\,\phi)
+ \O(\epsilon^2)\,, \nonumber \\
u_\e^\mathrm{in}(x) &=& 1 + \e \sum_{n=0}^\infty
R_n^{(1)}(\theta,\,\phi) + \O(\epsilon^2)\,, \nonumber
\end{eqnarray}
 % ------------- %
hence the continuity condition at the surface gives
 % ------------- %
\begin{equation}
 S_n^{(1)} = R_n^{(1)} + X_n\,.
\label{con1}
\end{equation}
 % ------------- %
As for the normal derivatives, in the first order we may consider
only the derivative in the radial direction, which gives
 % ------------- %
\begin{eqnarray}
\grad_r u_\e^\mathrm{ext}(x) &=& -1
+2 \e \sum_{n=0}^\infty X_n(\theta,\,\phi)
- \e \sum_{n=0}^\infty (n+1) S_n^{(1)}(\theta,\,\phi)
+ \O(\epsilon^2)\,, \nonumber \\
\grad_r u_\e^\mathrm{in}(x) &=& \e \sum_{n=0}^{\infty} n
R_n^{(1)}(\theta,\,\phi) + \O(\epsilon^2)\,, \nonumber
\end{eqnarray}
 % ------------- %
leading by comparison of the first-order terms to the condition
 % ------------- %
\begin{equation}
2 X_n - (n+1) S_n^{(1)} - n R_n^{(1)} = -R_n^{(1)} -
\alpha^{(1)}\,. \label{con2}
\end{equation}
 % ------------- %
Now the equations (\ref{con1}) and (\ref{con2}) yield
$\alpha^{(1)} = -X_0$ and $R_0^{(1)}= -X_0$; recall that by
convention we have $S_0^{(1)}= 0$. Furthermore, for $n \geq 1$ we
get
 % ------------- %
\begin{eqnarray}
S_n^{(1)} &=& \frac{1+n}{2n}X_n\,, \nonumber \\
R_n^{(1)} &=& \frac{1-n}{2n}X_n\,. \label{coef}
\end{eqnarray}
 % ------------- %
Next we employ Green's theorem by which we have
 % ------------- %
\begin{eqnarray}
\int_{\man A_\e^\mathrm{ext}} (\laplace u_\e)(x)\, \d^3 x &=&
-\int_{\man A_\e} \frac{\pd u_\e(x)}{\pd n_e}\, \d \nu_\e (x)
- 4\pi\,, \nonumber \\
 \int_{\man A_\e^\mathrm{in}} (\laplace u_\e)(x)\, \d^3x &=&
 -\int_{\man A_\e} \frac{\pd u_\e(x)}{\pd n_i}\,\d \nu_\e (x)\,, \nonumber
\end{eqnarray}
 % ------------- %
where the left-hand sides vanish by assumption. Summing then the
equations and using the boundary condition (\ref{BoundCon}) we get
 % ------------- %
$$
\alpha_\e \int_{\man A_\e} u_\e(x)\, \d \nu_\e(x) = 4 \pi\,.
$$
 % ------------- %
Substituting now the expansion (\ref{surel}), the above Ansatz up
to the second order and the coefficients (\ref{coef}) we
arrive at the condition
 % ------------- %
\begin{equation}
\alpha^{(2)} = X_0^2 - \frac{1}{4 \pi} \sum_{n=1}^\infty
\frac{1}{2}\left(n^2 + \frac{1}{n} \right) I_n\,, \label{crits}
\end{equation}
 % ------------- %
where we have denoted $I_n:=\int_{\man A_0} X_n^2(\theta,\,\phi)\,
\d \nu_0(\theta,\,\phi)$ and used the orthogonality of $X_n$
together with the known angular-momentum formula
 % ------------- %
$$
\int_{\man A_0} \left( \left(\frac{\pd X_n(\theta,\,\phi)}{\pd \phi}\right)^2
+ \frac{1}{\sin^2 \phi} \left(\frac{\pd X_n(\theta,\,\phi)}{\pd \theta}\right)^2 \right)
\sin \phi\, \d \phi \d \theta = n(n+1)I_n.
$$
 % ------------- %
Using the derived coefficients, we get an explicit asymptotic
formula for $\alpha_\e$,
%-------- %
$$
\alpha_\e = 1 -\e X_0 +\e^2 X_0^2 - \e^2 \frac{1}{4 \pi}
\sum_{n=1}^\infty \frac{1}{2}\left(n^2 + \frac{1}{n} \right) I_n +
\O(\epsilon^3)\,.
$$
On the other hand, there is a well known formula \cite[Sec.
1.33]{PolyaSzego} for the surface radius $\overline{S}_\e$, namely
 % ------------- %
$$
\overline{S}_\e = 1 + \e X_0 + \frac{1}{4 \pi} \e^2
\sum_{n=1}^{\infty} \left(\frac{n^2 +n +4}{4} \right) I_n +
\O(\epsilon^3)\,;
$$
 % ------------- %
combining these two expressions we get
 % ------------- %
$$
\alpha_\e \overline{S}_\e = 1 - \frac{1}{4 \pi} \e^2
\sum_{n=1}^{\infty} \left(\frac{1}{2}n^2 + \frac{1}{2n} -
\frac{n^2 +n +2}{4}\right)I_n + \O(\epsilon^3)\,.
$$
 % ------------- %
Since the $I_n$'s are non-negative, it is easy to see that
$\alpha_\e \overline{S}_\e \leq 1$, and moreover, that the
inequality is strict unless $I_1$ is the only nontrivial term,
i.e. $\rho(\theta,\,\phi) = X_1(\theta,\,\phi)$. To prove that
even in this case a small nontrivial deformation leads to
diminishing of the product $\alpha_\e \overline{S}_\e$ we need to
compute the next term in the asymptotic expansion, which means
here the fourth one because the third is zero. This can be done
explicitly by putting
 % ------------- %
$$
X_1(\theta,\,\phi) := A Y_0(\theta,\,\phi) +B
Y_{-1}(\theta,\,\phi) +C Y_1(\theta,\,\phi)\,,
$$
 % ------------- %
where $Y_i$ is the standard basis of the first-order spherical
harmonics, explicitly
 % ------------- %
\begin{eqnarray}
Y_0(\theta,\,\phi)&=&\sqrt{\frac{3}{4 \pi}}\, \cos \phi\,, \nonumber \\
Y_1(\theta,\,\phi)&=&\sqrt{\frac{3}{4 \pi}}, \sin \phi \cos \theta\,, \nonumber \\
Y_{-1}(\theta,\,\phi)&=&\sqrt{\frac{3}{4 \pi}}\, \sin \phi \sin
\theta \nonumber\,.
\end{eqnarray}
 % ------------- %
After a lengthy but tractable computation we arrive at the
expression
 % ------------- %
$$
\alpha_\e \overline{S}_\e =1 - \e^4 \frac{3(A^2 + B^2 +C^2)^2}{20
\pi} + \O(\epsilon^5)\,,
$$
 % ------------- %
which proves the desired claim.

\setcounter{equation}{0}
%%%%%%%%%%%%%%%%%%%%%%%%%%%%%%%%%%%%%%%%%%%%%%%%%%
\section{A large deformation example}
\label{sec:global}

The aim of this section is to show that the local result of
Theorem~\ref{def:local} does not extend to general
surface-preserving deformations: we are going to show that for any
fixed interaction strength $\alpha_0$ there is a surface $\man A$
of unit area such that a constant singular interaction, $\alpha(x)
:= \alpha_0$, does not induce existence of bound states. The
example idea is to construct a surface with a large diameter, that
is, to examine the situation when the capacity is much greater
than the surface radius.

The way to achieve this goal is to show that there are surfaces
for which the strict inequality
%---------%
$$
\|(\alpha \Gamma)^2\| < \|\alpha\Gamma\|
$$
%---------%
holds, where $\Gamma$ is the Birman-Schwinger-type operator from
Proposition~\ref{def:crucial}. Since $\alpha\Gamma$ is strictly
positive by assumption, this will yield the inequality $\|\alpha
\Gamma\| <1$ implying by the said proposition that the
corresponding operator $\opr{H}_\alpha$ has no bound states.

We have demonstrated that $\|\alpha \Gamma\|$ is an eigenvalue of
the operator $\alpha \Gamma$ corresponding to a positive
eigenfunction which, in particular, implies
%----------%
$$
\|\alpha \Gamma\| = \sup_{f \in L^2(\man A,\d\nu),\,f > 0}
\frac{\|\alpha \Gamma f\|}{\|f\|}\,;
$$
%----------%
note that the supremum is taken over positive functions only. Next
we employ a simple geometric inequality
%---------%
\begin{eqnarray*}
\frac{1}{|x-y||y-z|} &=& \left(\frac{1}{|x-y|}
+\frac{1}{|y-z|}\right)\frac{1}{|x-y|+ |y-z|}
\\ &\leq& \left(\frac{1}{|x-y|} +\frac{1}{|y-z|}\right)
\frac{1}{|x-z|}
\end{eqnarray*}
%--------%
which allows us to estimate
%----------%
\begin{multline}
((\alpha \Gamma)^2f)(x) = \alpha_0^2 \frac{1}{(4 \pi)^2} \int_{A
\times A} \frac{1}{|x-y|}\frac{1}{|y-z|}\, f(z)\,
\d \nu(y) \d \nu(z) \\ \leq \frac{\alpha_0^2}{(4 \pi)^2}
\int_{A \times A} \left(\frac{1}{|x-y|} +\frac{1}{|y-z|}\right)\,
\frac{1}{|x-z|}\, f(z)\, \d \nu(y) \d \nu(z) \\
\leq 2 \frac{\alpha_0^2}{(4 \pi)}\, \|\Gamma\|_\infty \int_{A}
\frac{1}{|x-z|}\, f(z)\, \d \nu(z) = 2 \alpha_0 \|\Gamma\|_\infty
(\alpha \Gamma f)(x), \nonumber
\end{multline}
%---------%
for any $f$ nonnegative, where we have put
%---------%
$$
\|\Gamma\|_\infty := \frac{1}{4 \pi} \sup_{x \in \man A} \int_\man
A \frac{1}{|x-y|}\, \d \nu(y)\,;
$$
%--------%
the integral obviously converges for any $x\in\man A$ and is
continuous w.r.t. this variable, so $\|\Gamma\|_\infty$ is finite.
By taking the supremum  over all positive functions $f \in
L^2(\man A,\d\nu)$ we then get
%--------%
\begin{equation}
\|(\alpha \Gamma)^2\| \leq 2 \alpha_0 \|\Gamma\|_\infty \|\alpha
\Gamma\|\,. \label{kvadrat}
\end{equation}
%-------%
Now we will show that the quantity $\|\Gamma\|_\infty$ tends to
zero with the increasing diameter of the surface $\man A$. To do
this we choose a smooth \emph{positive} function
 % ------------- %
$$f : (-1,\,1) \to \mathbb{R}\,, \quad
2 \pi \int_{-1}^{1} f(v) \d v =1\,, $$
 % ------------- %
such that $f(\pm 1)=0$ and $f'(\pm 1)=\pm\infty$ and define a
family of surfaces $\man A_\epsilon$ by the equations
%----------%
\begin{eqnarray}
x &=& \epsilon f(v) \cos u\,, \nonumber \\
y &=& \epsilon f(v) \sin u\,, \nonumber \\
z &=&  \epsilon^{-1} v\, \nonumber,
\end{eqnarray}
%----------%
where $v \in (-1,\,1),\,u \in (0,\,2 \pi)$.
Thus we have
%---------%
$$
\d \nu_\e(u,\,v) = f(v) \sqrt{1 +\e^4 f'(v)^2}\,;
$$
%---------%
as a consequence, the surface area satisfies $S_\epsilon >1$ for
all $\e>0$ and approaches one in the limit $\e \to 0$. On the
other hand for the quantity $\|\Gamma_\e\|_\infty$ we have
%----------%
\begin{multline}
4 \pi \|\Gamma_\e\|_\infty =
\sup_{x,y} \int_{-1}^1 \d v \int_0^{2 \pi} \d u \\
\frac{f(v) \sqrt{1 + \e^4 f'(v)^2}}{\sqrt{\e^2(f(v) \cos u - f(x)
\cos y)^2 + \e^2(f(v) \sin u - f(x) \sin y)^2 + \e^{-2}(v-x)^2}}
\\\leq \sup_x \int_{-1}^1 \d v \frac{M}{\sqrt{N\e^2  + \e^{-2} (v-x)^2
}}\,, \nonumber
\end{multline}
%---------%
for suitable, sufficiently larger constants $M,\,N$ which depend
on $f$ only. It is easy to see that the last integral is maximized
for $x=0$ giving
%--------%
$$
\|\Gamma_\e\|_\infty \leq 2M\epsilon\,
\mathrm{arcsinh}(\sqrt{N}\e^2)^{-1}
$$
%--------%
and since the right-hand side behaves as $-4M\epsilon \ln\epsilon
+ \O(\epsilon)$, we conclude that $\|\Gamma_\e\|_\infty \to 0$ as
$\e \to 0$. It is therefore possible to choose $\e$ in such a way
that $2 \alpha_0 \| \Gamma_\e\|_\infty <1$ which by means of
(\ref{kvadrat}) implies $\|(\alpha \Gamma_\e)^2\| < \|\alpha
\Gamma_\e\|$, and consequently, the operator $\opr{H}_\alpha$
corresponding to elongated enough surface $\man A_\epsilon$ has
empty discrete spectrum.

%%%%%%%%%%%%%%%%%%%%%%%%%%%%%%%%%%%%%%%%%%%%%%%%%%
\subsection*{Acknowledgments}
The research was supported in part by the Czech Ministry of
Education, Youth and Sports within the project LC06002.

%%%%%%%%%%%%%%%%%%%%%%%%%%%%%%%%%%%%%%%%%%%%%%%%%%

\end{document}